# Interferometric carrier-envelope phase stabilization for ultrashort pulses in the mid-infrared

M. Meierhofer, S. Maier, D. Afanasiev, J. Freudenstein, C. P. Schmid, R. Huber

*Department of Physics, University of Regensburg, 93040 Regensburg, Germany*

**We demonstrate an active carrier-envelope phase (CEP) stabilization scheme for optical waveforms generated by difference-frequency mixing of two spectrally detuned and phase-correlated pulses. By performing ellipsometry with spectrally overlapping parts of two co-propagating near-infrared generation pulse trains, we stabilize their relative timing to 18 as. Consequently, we can lock the CEP of the generated mid-infrared (MIR) pulses with a remaining phase jitter below 30 mrad. Employing these pulses for high-harmonic generation in a bulk semiconductor validates our technique. This experiment reveals that our method also stabilizes the energy of the MIR pulses, thereby approaching the intrinsic stability of the underlying laser system.**

When the duration of ultrashort light pulses approaches the few-cycle regime, the carrier-envelope phase (CEP), defined as the phase between the carrier wave and the pulse envelope, starts to gain importance [1]. After the first observation of a CEP-dependent effect in the photoionization of krypton gas with a yet unstabilized laser system [2], the invention of *f-2f* interferometry allowed for a reliable measurement and stabilization of the CEP of visible and near-infrared light pulses [3]. However, in the mid-infrared (MIR), where CEP-dependent phenomena have, for example, been observed in high-harmonic generation (HHG) from solids [4-7], detecting the CEP by *f-2f* interferometry is not practical.

In an increasingly popular scheme, ultrashort MIR pulses have been obtained by difference-frequency generation (DFG) of two spectrally detuned but phase-correlated near-infrared pulses interacting in a $\chi^2$-nonlinear medium [8]. In principle, this process results in a stable CEP, as the phases of the two generation pulses cancel [9]. However, changes in the environmental conditions, such as temperature drift or air turbulences, can alter the optical path length of the two generation pulses, thereby introducing an additional phase difference. By difference-frequency mixing, this phase difference directly translates into drifts of the CEP of the MIR pulses. To tackle this challenge, only few active stabilization techniques have been proposed: Manzoni *et al.* have employed the interference of an auxiliary gate pulse and a frequency-shifted copy of it, which is generated by DFG with the original MIR pulse [10]. Liu *et al.* have used the interference between spectrally overlapping parts of the two generation pulses [11], while Yamakawa *et al.* have traced the CEP of MIR pulses directly in the time domain by simultaneous electro-optic sampling [12]. Despite the success of these approaches, there are challenges and limitations, such as the slow readout of spectrometers, the need for additional pulse energy, or uncorrelated drift in the stabilization setup itself.

In this letter, we introduce a fast photodiode-based CEP stabilization scheme. Its compact common-path geometry renders the setup drift-free. Furthermore, only one-thousandth of the original pulse energy is required for the stabilization. By analyzing the polarization state of the spectral overlap of the interfering generation pulses, we measure their relative time delay. A PID controller stabilizes the timing jitter and consequently the CEP of the generated MIR waveforms. We experimentally benchmark the performance of our technique by comparing high-harmonic generation in the bulk semiconductor gallium selenide (GaSe) with and without CEP stabilization [4].

Figure 1a shows the experimental setup [13]. A Ti:sapphire laser amplifier provides pulses with a duration of 33 fs and an energy of 5.5 mJ at a repetition rate of 3 kHz. These pulses are employed to pump a dual-branch optical parametric amplifier (OPA), which generates two spectrally detuned signal pulse trains with center wavelengths of 1.2 μm (OPA A) and 1.33 μm (OPA B), respectively. To guarantee a mutual phase relation, both arms are seeded by the same white-light continuum. The two orthogonally polarized pulse trains are spatially overlapped by a dichroic mirror and sent onto a GaSe crystal to generate MIR pulses by type-II DFG. The temporal overlap, $t$, between the two OPA pulses is controlled with a fused silica wedge pair. As the phase of the generated MIR pulses depends on the relative phase, $\Delta\varphi$, between the two generation pulses ($\varphi_{MIR} = -\pi/2 + \Delta\varphi$) [9], the CEP can be adjusted by changing the delay between them. A germanium wafer is employed as a long-pass filter to separate the pump light from the MIR pulses. The latter are subsequently reflected off an indium-tin-oxide-coated window, where they are spatially overlapped with ultrashort gate pulses for electro-optic sampling (EOS) [14]. To generate the gate pulses, a small fraction of the output of OPA A is focused into a YAG window for white-light continuum generation. The spectral region from 740 to 940 nm is subsequently compressed by a pair of fused silica prisms to a pulse duration of 8 fs. The EOS delay time, $\tau$, is scanned using another wedge pair.

As the two signal pulse trains travel on separate beam paths, their relative timing can drift over time, leading to significant changes in the CEP of the generated MIR pulses. To monitor the relative delay between them, we perform ellipsometry of spectrally overlapping parts of the two co-propagating generation pulses. The setup for this is highlighted in Fig. 1a with a dashed box. An uncoated fused silica plate (FSP) picks off 3% of the energy of the spatially overlapped NIR pulses, which are subsequently attenuated by a variable neutral density filter wheel (ND) to below 3 mW, corresponding to only one per mille of the incoming average power. This is necessary to prevent the detection setup from saturating. A 10 nm bandpass centered at 1250 nm selects the spectral overlap of the two near-infrared pulses. Due to their crossed polarization, the polarization of the superimposed pulses can continuously vary between linear and circular only depending on the relative delay between the two OPA arms.

By setting the fast axis of a quarter-wave plate ($\lambda/4$) to 45° with respect to the fundamental pulses, the polarization state of the superimposed pulses can be analyzed [15]: If both pulses oscillate in phase, the resulting polarization is linear, which is not altered by the quarter-wave plate (see Fig. 1c). Accordingly, a Wollaston prism splits the pulses equally, and the two photodiodes are balanced. If there is a phase difference of ±90° between the two pulses, the resulting field is either left- or right-circularly polarized (see Fig. 1b,d), which is transformed to horizontal or vertical polarization by the quarter-wave plate, respectively. Therefore, the Wollaston prism ensures that only one of the two photodiodes is illuminated. Hence, the differential photocurrent of the two diodes provides a sinusoidal feedback signal (Fig. 1a, inset). In the following, we use this signal to trace and stabilize the relative delay of the two generation pulses and, consequently, the CEP of the MIR pulses.

By adjusting the wedge position, the relative delay between the two OPA arms is set to zero. There, the feedback signal crosses zero (see Fig. 1a, inset), which allows us to monitor the temporal drift between the two pulse trains. Figure 2a shows the evolution of the relative delay over 20 minutes. Without active stabilization, the overlap drifts by approximately 50 as per minute while exhibiting RMS fluctuations of 191 as. If instead, we feed the signal to a PID controller, which adjusts the wedge position accordingly, we can eliminate the drift and reduce the fluctuations by more than one order of magnitude to 18 as. This translates to a CEP stability of the generated MIR pulses of 300 mrad and 28 mrad without and with feedback, respectively. The improvement in the stability can be visualized when plotting a histogram of the relative delays (Fig. 2b). In the free-running case (blue), the distribution is not centered around zero, indicating long-term drifts, and features a broad bandwidth due to large timing fluctuations. In contrast, the feedback loop fully eliminates drift and compensates for fluctuations, resulting in a narrow and centered histogram (orange). Figure 2c shows the Fourier transforms of the time-domain signals in Fig. 2a. With activated feedback (orange), the low-frequency components below 0.1 Hz are suppressed by up to three orders of magnitude, as compared to the unstabilized case (blue). This confirms that slow processes, such as temperature drift or air turbulences, can be fully compensated for by our stabilization scheme.

Figure 3a shows exemplary field transients of two actively CEP-stabilized MIR pulses with a center frequency of 25 THz measured via EOS in a 6.5-µm-thick ZnTe crystal. By using the falling or rising edge of the sinusoidal feedback signal, we can lock the CEP to two values, which are separated by $\pi$. In this case, we locked the feedback loop close to - $\pi/2$ (orange) and $\pi/2$ (blue), respectively. Yet the strongly suppressed timing jitter allows the PID controller to stably work not only at the zero crossings. Almost the entire slopes can be used to actively control the CEP. Figure 3b demonstrates the range of the accessible CEPs. By adjusting the setpoint of the PID controller to different delays, we lock the phase of the MIR pulses to various values. For each setpoint, an EOS transient is recorded, and from each time-domain trace, we extract the CEP by Fourier transform (black dots). Here, we account for a constant offset due to the Gouy phase shift in the detection focus. The black dashed line represents the theoretically expected dependence of the CEP on the delay with a slope of $2\pi/\lambda_B = 2\pi/1.2$ µm.

We find stable locking for a large part of both slopes, covering two intervals of a width of $\pi/2$ each. This directly reflects the sinusoidal shape of the feedback signal, where stabilization is not possible in close vicinity of the minima and maxima (see Fig. 1a, inset). Note that delays, where no stabilization is possible in this configuration, can easily be accessed by replacing the quarter-wave plate with a half-wave plate at a 22.5° angle. The resulting feedback signal is shifted by π/2, creating slopes where the minima and maxima are located for the quarter-wave plate.

The center frequency of the MIR pulses can be readily tuned by rotating the GaSe emitter crystal, hence changing the phase-matching condition. Tuning the MIR center frequency leaves the setup for CEP control unaffected since the light for the stabilization is split off before the generation crystal. Therefore, our scheme naturally allows for stabilizing the CEP of pulses with variable center frequencies.

To prove the experimental impact of the CEP stabilization scheme, we employ pulses centered around 29 THz (peak field strength, ~30 MV/cm) to drive high-harmonic generation in a bulk GaSe crystal (thickness, 70 μm). In this process, the MIR electric field drives extremely non-resonant interband polarization combined with coherent intraband currents resulting in dynamical Bloch oscillations. These dynamics manifest in the emission of ultrabroadband radiation covering more than 12 optical octaves [4]. Due to the non-perturbative nature of the underlying strong-field light-matter interaction, HHG is highly sensitive to the exact driving waveform [13].

Figure 4a displays normalized high-harmonic spectra collected over a measurement time of 1.5 h. In the first half of the acquisition time, the system is free running, while in the second half, we activate the CEP lock. To quantify the increase in stability, we fit every harmonic order with a Gaussian function and extract the center wavelengths and the peak intensities. The center wavelengths are displayed in Fig. 4a as white lines. The peak positions of all observed harmonic orders become more stable when the feedback is activated. An evaluation of the center wavelengths of the 12$^{th}$ harmonic, centered around 876 nm, reveals a 57% reduction of fluctuations (standard deviation) of the spectral position. This directly proves that our stabilization technique works.

In Fig. 4b, we depict exemplarily the extracted peak intensity of the 12$^{th}$ harmonic order, $I_{HH12}$. Here, we find a 77% reduction of intensity fluctuations associated with the harmonic peaks, with remaining fluctuations as low as 6%. The intensities of all other harmonics follow a qualitatively similar behavior. As HHG is highly sensitive to the driving electric field strength [4], variations in the MIR pulse energy translate to intensity changes of the emitted harmonics. Our CEP stabilization technique relies on stabilizing the relative delay between the two generation pulse trains. Since the generation pulses have a duration of only 50 fs, our method hence stabilizes the MIR pulse energy as well.

It is noteworthy that the regulation bandwidth is currently limited to 10 Hz by the lock-in amplifier and the software-based PID controller used here. By replacing both with boxcar integration and a PID in real-time electronics, single-shot detection and control could be achieved, even at repetition rates in the MHz range. Yet more intriguingly, we envision that our presented stabilization technique may also be applied whenever the timing of two orthogonally polarized pulses with spectral overlap is to be stabilized. This is particularly interesting for DFG-based sources such as OPAs or OPCPAs. The balanced detection scheme could replace single-diode detection or spectrometer-based setups, thus allowing for faster feedback with a higher signal-to-background ratio.

In conclusion, we have demonstrated a straightforward and powerful method to actively stabilize and control ultrashort pulses in the MIR generated by DFG. By performing ellipsometry on the spectral overlap of the two generation pulse trains, we obtained a sinusoidal feedback signal as a function of the time delay between the two. The slopes of this signal allowed us to stabilize the temporal overlap between the two pulses with a precision of 18 as. Thus, we successfully control the CEP of the generated MIR pulses and stabilize their pulse energy. Finally, we utilized the actively stabilized MIR pulses to drive HHG in GaSe. By measuring the high-harmonic spectra over hours, we managed to directly showcase the effects of the stabilization in the high-harmonic center wavelengths and intensities.

In a first application campaign, we have recently employed this stabilization technique to access Coulomb correlations between Bloch electrons in $WSe_2$ directly in the time domain [16]. We anticipate that our method will also have a strong impact on the fidelity of lightwave electronic [17], valleytronic [18], or plasmonic [19] devices, where the magnitude and phase of light pulses have to precisely steer the motion of electrons. Additionally, this approach will be applicable to lightwave STM [20], where the exact shape of the electric field governs the time window for tunneling of single electrons.

**Funding.** This work has been supported by the Deutsche Forschungsgemeinschaft (DFG, German Research Foundation) through Project ID 422 314695032-SFB 1277 (Subproject A05) as well as Research Grant HU1598/8.

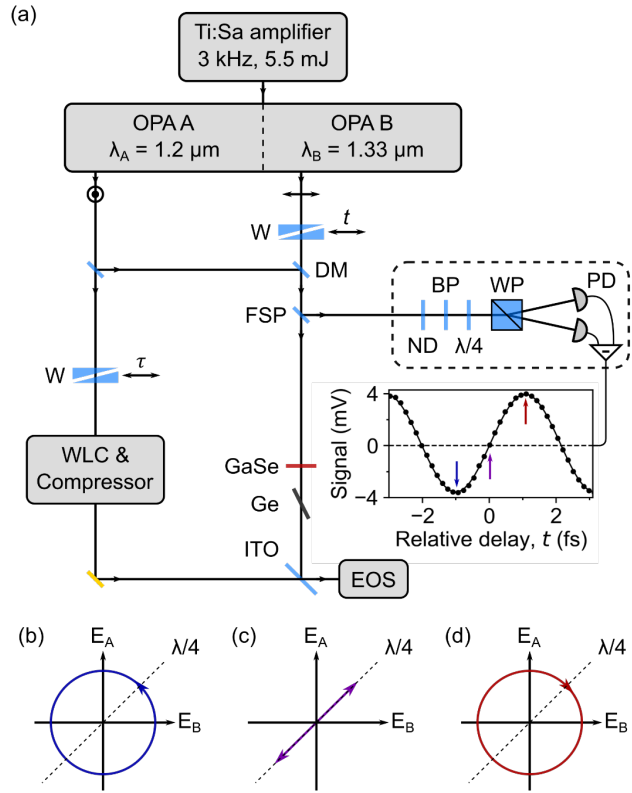

**Figure 1 a)** Experimental setup for generation and detection of actively CEP stabilized MIR pulses. OPA, optical-parametric amplifier; W, wedge pair; DM, dichroic mirror; FSP, fused silica plate; ND, neutral density filter; BP, 10 nm bandpass filter centered around 1250 nm; λ/4, quarter-wave plate; WP, Wollaston prism; PD, photodiode; GaSe, gallium selenide crystal; Ge, germanium wafer; ITO, indium-tin-oxide-coated window; WLC & Compressor, white-light continuum and compression; EOS, electro-optic sampling. **b)-d)** Possible polarization states of the light pulses impinging on the quarter-wave plate, whose fast axis is indicated by the dashed line. By analyzing this polarization state with a quarter-wave plate and a Wollaston prism, a sinusoidal feedback signal (Fig. 1a, inset) is generated. We use its monotonic regions to measure and stabilize the temporal overlap between the two OPA pulses, which in turn allows for interferometrically precise CEP stabilization.

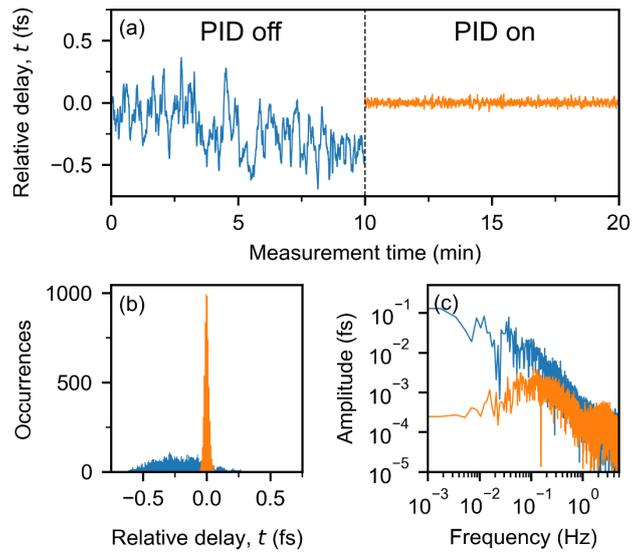

**Figure 2 a)** Relative delay between the two OPA pulse trains over a measurement time of 20 minutes. Without feedback (blue), the temporal overlap drifts with approximately 50 as/min while exhibiting RMS fluctuations of 191 as. When the feedback is activated (orange), the drift is eliminated, and the remaining fluctuations are reduced to 18 as. This corresponds to a CEP stability of the generated MIR pulses of 300 mrad and 28 mrad, respectively. **b)** Histograms of the relative time delays in a) with feedback off (blue) and on (orange). **c)** Fourier transform of the time-domain signal in a). With feedback applied (orange), the low-frequency components are suppressed by up to three orders of magnitude compared to the unstabilized case (blue).

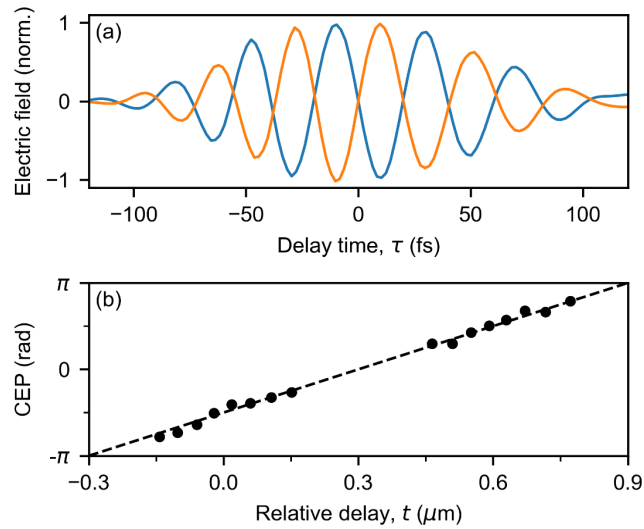

**Figure 3: CEP control. a)** Electric-field transients of two CEP-stabilized MIR pulses with a center frequency of 25 THz for a CEP of - π/2 (orange) and π/2 (blue). **b)** Extracted CEP values (black dots) for different delays. The strongly reduced CEP fluctuations allow us to not only stabilize at the zero crossings of the feedback signal but use both slopes to control the CEP. The black dashed line represents the theoretically expected dependence of the CEP on the delay with a slope of 2π/1.2 µm.

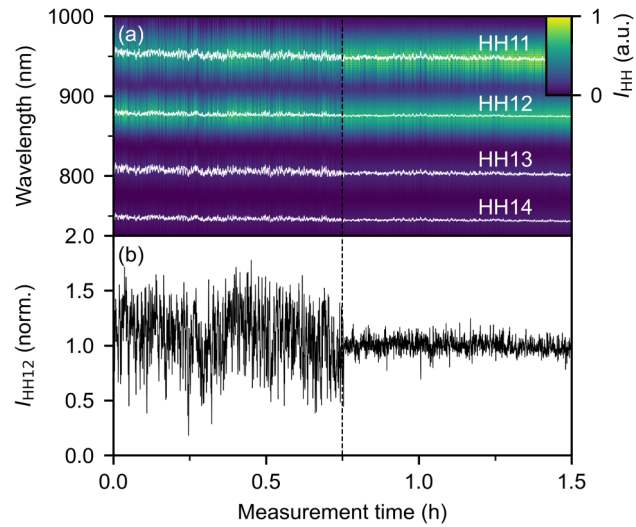

**Figure 4: Influence of CEP stabilization on high-harmonic generation. a)** Normalized high-harmonic spectra from GaSe driven by 29 THz pulses over 1.5 h. The peaks of each harmonic order are marked by a white line. As the spectral position is sensitive to the CEP, the increased stability directly proves the reduction of CEP noise by our stabilization. **b)** Extracted intensity of the 12$^{th}$ harmonic order, $I_{HH12}$, centered around 876 nm. The intensity fluctuations are reduced by 77%. This indicates that the stabilization scheme stabilizes not only the CEP but also the pulse energy of the MIR pulses.